\documentclass[12pt,preprint]{aastex}

\shorttitle{Vela X at 31 GHz} \shortauthors{Hales et al.}

\begin{document}


\title{ Vela X at 31 GHz}



\author{Hales A. S.\altaffilmark{1} , Casassus S. , Alvarez H. , May J. \& Bronfman L.}

\affil{Departamento de Astronom\'{\i}a, Universidad de Chile,
    Casilla 36-D, Santiago, Chile.}

\author{Readhead A. C. \&  Pearson T. J.} 
\affil{105-24 California Institute of Technology, Pasadena, CA
91125}

\author{Mason B. S.}
\affil{National Radio Astronomy Obseratory, Green Bank, WV 24944}


\and

\author{Dodson R.}
\affil{University of Tasmania, Department of Mathematics and
Physics, GPO Box 252-21, Hobart Tasmania 7001 Australia.}

\altaffiltext{1}{Present address: Department of Physics and Astronomy,
University College London, Gower street, London, WC1E 6BT}





\begin{abstract}

We present observations of the Vela X region at 31 GHz using the
Cosmic Background Imager (CBI). We find a strong compact radio source
($5.9'\times 4.1'$, FWHM) about the Vela pulsar, which we associate
with the Vela pulsar wind nebula (PWN) recently discovered at lower
radio-frequencies. The CBI's 4$'$ resolution for a 45$'$ field of view
allows the PWN to be studied in the large-scale context of Vela~X.
Filamentary structure in Vela X, which stands out in lower frequency
maps, is very low-level at 31~GHz. By combining the 10 CBI channels,
which cover 26-36~GHz, and 8.4 GHz archive data, we study the spectral
energy distribution (SED) of the PWN and the brightest filaments. Our
results show that the spectral index $\alpha$ ($F_{\nu}\propto
\nu^\alpha$) of the PWN is flat, or even marginally positive, with a
value of $\alpha_{8.4}^{31} = 0.10\pm0.06$, while the Vela X
filamentary structure has a negative spectral index of
$\alpha_{8.4}^{31} = -0.28\pm0.09$. The SED inhomogeneity observed in
Vela X suggests different excitation processes between the PWN and the
filaments. We investigate whether the PWN's flat spectrum is a
consequence of variability or truly reflects the SED of the object. We
also investigate the nature of the Vela X filamentary structure. A
faint filament crosses the PWN with its tangent sharing the same
position angle as the PWN major axis, suggesting that it might be an
extension of the PWN itself. The SED and bulk morphology of Vela X are
similar to those of other well-studied plerions, suggesting that it
might be powered by the pulsar. The peak of the PWN at 31~GHz is
$80''\pm20''$ south-west of the peak at 8.4~GHz. This shift is
confirmed by comparing the 31 GHz CBI image with higher resolution
5~GHz Australia Telescope Compact Array observations, and is likely to
be due to SED variations within the PWN.


\end{abstract}




\section{Introduction}


When studied with low resolution at radio frequencies, the Vela
supernova remnant(SNR) has usually been divided into three different
regions of enhanced emission, Vela X, Y and Z (see
\citet{bo98}). While the Vela Y and Z components have always been
unambiguously associated with the SNR's shell, there has been much
controversy about the nature of Vela X, the brightest
component. Observations of the Vela X region have shown that the
0.84--8.4~GHz emission is concentrated in a network of radio
filaments, superimposed on a smooth plateau of diffuse emission
\citep[and references therein]{mi95}. In the high resolution (43$''$)
MOST mosaic of the complete Vela remnant built at 0.843~GHz by
\citet{bo98}, the filaments make up the bulk of the high spatial
frequency emission.  Based on its filamentary morphology and/or on
spectral energy distribution (SED) differences with Vela Y and Z, a
group of authors \citep{we80,dw91,bo98,bo02,fr97b,al01} have supported
the hypothesis that Vela X is the pulsar-powered (plerionic) core of
the Vela SNR.  By contrast, \citet{mi86} have argued that the Vela X
spectral index is not very different from that of the rest of the
remnant, concluding that Vela X ``is not directly pulsar-driven but is
an enhanced region of shell emission''.


Different hypotheses for the nature of Vela X have been
proposed. Comparison between 8.4~GHz and 2.7~GHz single-dish
observations led \citet{mi95} to conclude that ``the spectral index is
remarkably constant (...)  over the brightest parts of the SNR'', with
a value of about $\alpha \sim -0.4$ to $-0.8$ ( in this work we use
$\alpha$ to refer to the spectral index of the spectral energy
distribution in Jy, $F\propto\nu^{\alpha}$) . This result allowed him
to argue that in Vela X there is no evidence for radial steepening of
the spectral index with increasing distance to the pulsar, as seen in
the Crab and other pulsar powered remnants \citep{ve92}), supporting
his hypothesis of a non-pulsar powered Vela X.

However, morphological and spectral differences between Vela X, Y and
Z and the off-center position of the Vela pulsar have encouraged the
development of new models for the nature of the Vela X filamentary
structure. \citet{rey84} proposed that the filamentary structure and
the shifting of the bulk of Vela X from the pulsar's position are the
result of the crushing of the original Vela pulsar wind nebula (PWN)
by a reverse shock wave, produced by the interaction between the SNR
and its surrounding medium. As modeled in detail by \citet{bl01},
inhomogeneities in the surrounding medium would create an asymmetric
reverse shock wave that would compress the primitive PWN, displacing
it from its original position, thus explaining the asymmetric
appearance of Vela X and the off-center position of the pulsar. An
alternative model has been proposed by \citet{ma95} after studying the
one sided X-ray `jet' feature extending 40$'$ south of the pulsar
position. They suggested a pulsar-powered nature for this X-ray `jet',
which was subsequently found to have a bright radio counterpart
\citep{mi95,fr97b}.



Filled-center components, or ``plerions'', in the cores of supernova
remnants are taken as evidence that pulsars deposit energy into their
surroundings \citep{fr97,rei98}. Plerions are characterized by a flat
radio spectrum, high degrees of linear polarization, and present
irregular morphologies. They are currently understood as driven by the
slow-down of the exciting neutron star, whose kinetic energy loss rate
can account for the nebular radiative losses seen in their wind nebula
\citep{re74,ke84,ra01,ch03}. However only about $15\%$ of the entries
in the SNR catalog\footnote{\tt
http://www.mrao.cam.ac.uk/surveys/snrs} by \citet{gr04} are classified
as plerions or combined-type SNRs, a fraction much smaller than
expected for core collapse events in our galaxy
\citep{rei98}. Galactic surveys with the VLA \citep{fr97} and with the
Effelsberg 100-m telescope \citep{rei02} have tried to raise this
fraction, but in spite of all efforts the number of known plerions is
still small. \citet{rei98} has suggested that it is possible that ``a
number of shell-type SNRs are misclassified in that they have a weak
central component which, when seen against the dominating steep
spectrum shell emission, is hidden at low frequencies''. He proposes
to combine both low and high-frequency observations, in order to
unveil the flat spectrum cores. \citet{rei02} also proposes that the
spatial-frequency decomposition of these cores will permit the study
of their small-scale structures, which are thought to provide hints on
how the pulsar injects energetic particles in the surrounding nebula.



VLA observations of the immediate vicinity of the Vela pulsar by
\citet{bi91} discovered a `localized ridge of highly-polarized radio
emission' located North-East of the Vela pulsar. Using scaled array
VLA observations they derived a 1.4/4.8~GHz spectral index for this
feature much steeper than for any other previously studied structure
in Vela X. Based on its physical and geometric properties,
\citet{bi91} argued that this feature `is directly related to the Vela
pulsar', and compared it to the radio `wisps' seen in the Crab
nebula. Recent Australia Telescope Compact Array (ATCA) polarimetric
observations at 1.4 to 8.5~GHz \citep{le01,do03a} have confirmed the
pulsar-powered nature of this nebula surrounding the pulsar. Using
combined array configurations, \citet{do03a} identified a symmetric
nebula around the projected pulsar spin axis, which they associated to
the Vela pulsar as its PWN. Their polarimetric data also revealed a
toroidal magnetic field structure similar to that of the compact X-ray
nebula surrounding the pulsar \citep{pav01a,pav01b}. Interestingly,
the NE lobe of the PWN is markedly brighter in peak radiation
intensity at 5~GHz than the SW lobe.

In this work we present the highest radio-frequency map of Vela X to
date, which we obtained with the Cosmic Background Imager (CBI) radio
interferometer, whose 10 channels allow imaging the object in the
unexplored frequency range of 26 to 36~GHz. We used the CBI's high
sensitivity for a wide instantaneous field of view (45$'$, FWHM) in
order to study the PWN in the large-scale context of Vela~X, and to
quantify spectral index variations. Comparison of the 31~GHz image
with 8.4~GHz and 4.8~GHz archive data shows that the compact region
surrounding the Vela pulsar is markedly brighter relative to the Vela
X filaments at 31~Ghz than at lower frequencies. This result might
arise from SED differences between the compact source and the rest of
Vela X filaments structure, from variability of the source after the
January 2000 glitch (Milne's 8.4~GHz data were acquired in 1992), or
perhaps be a combination of both effects. To obtain comparable images
at other radio frequencies, we simulated the CBI interferometric
observations on an image at 8.4~GHz, which was kindly made available
to us by D.K. Milne and P. Reich, and the Vela X data from the
Parkes-MIT-NRAO survey \citep[PMN]{co93} at 4.8~GHz.


Our aim is to present for the first time a quantitative report on the
spectral differences between the Vela PWN and filaments. $\S2$ reports
on the observations and on the data reduction; $\S3$ describes image
analysis: Improvement of the 8.4~GHz Parkes image, simulation of CBI
observations over the Parkes and PMN data, as well as deconvolution
techniques. Results are discussed in $\S4$; $\S5$ summarizes our
conclusions.

\section{Observations and data reduction}

The Cosmic Background Imager (CBI, \citet{pa02}) is a 13-element
close-packed interferometer mounted on a 6-meter platform, with
baselines ranging from 1 to 5.5~m. It operates in ten 1-GHz frequency
bands from 26~GHz to 36~GHz. The array configuration for the Vela X
observations yielded a synthetic beam-width of $4.1'$ resolution over
an instantaneous field of view of $45'$, at full-width at half maximum
(FWHM).

The observations of Vela X were carried out in 2000 November, 2001
April and May, and 2003 April. We Nyquist-sampled a region of Vela X
using 4 pointings, spaced by $\sim20'$. The pointings's coordinates
($\alpha,\delta$, J2000) were: (08:35:20,$-$45:10:20),
(08:35:10,$-$45:28:00), (08:34:52,$-$45:48:00),
(08:32:30,$-$45:45:60), corresponding to the region surrounding the
Vela pulsar, to the middle and to the southern end of the radio
filament lying South of the pulsar \citep{fr97b}, and to another
bright filament, located approximately $50'$ to the South-West of the
pulsar. Integration times were 19639, 11478, 9347 and 6251 seconds,
respectively.  The data acquired were calibrated against Tau A using a
special-purpose package (CBICAL, T.J. Pearson). Details of the
calibration procedure are given in \citet{ma03}. No variability
between different observation dates was detected, so data from the
same pointing but different observation dates were merged.

Spill-over affects the lowest $(u,v)$ frequencies, which we cut out by
restricting the range in $uv$-radius to above 150~$\lambda$, losing
20\% of the data.

\section{Image analysis}

\subsection{Deconvolution and mosaicking of the CBI data}

In Vela X extended structure fills the primary beam, making
CLEAN-based deconvolutions very sensitive to the location of CLEAN
boxes.  We thus preferred deconvolving with the Maximum Entropy Method
(MEM), as it allows an unbiased comparison with maps at different
frequencies. We used the MEM algorithm implemented in the AIPS++
(\footnote{\tt http://aips2.nrao.edu/docs/aips++.html}) data reduction and
analysis package. Algorithm parameters were set in order to model
emission as deep as the theoretical noise expected from the system
sensitivity and integration times ($\sim5$~mJy~beam$^{-1}$). Uniform weights were
used in order to improve resolution.

After deconvolution, we linearly mosaicked the restored images using the
Perl Data Language (PDL\footnote{\tt http://pdl.perl.org}). Using all 10
CBI channels we obtained a 31~GHz image of Vela X, shown in Figure
\ref{fig:2clr}-a. Repeating the same procedure for the first and last
5 CBI channels, we also produced 33.5~GHz and 28.5~GHz mosaics of Vela X.

\subsection{Simulation of the CBI (u,v) coverage on the comparison  images}

We are interested in computing the spectral index between 8.4 and
31~GHz, by comparing the 31-GHz CBI image with the 8.4-GHz Parkes
image from \citet{mi95}. In Vela X, a network of filaments is
superimposed over a smooth plateau, to which the CBI is not sensitive
as it lacks total power measurements. Thus, comparison between CBI
31~GHz and single dish 8.4~GHz data is not direct. We simulated the
31-GHz CBI observations over the 8.4-GHz single-dish data, using the
MockCBI routine (T.J. Pearson). MockCBI simulates visibilities of any
given sky image matching the $uv$-coverage of a reference CBI
observation. We therefore produced 4 simulated CBI pointings of Vela X
over the 8.4-GHz data, using the identical pointing coordinates and
$uv$-coverage than of our original 31~GHz CBI observations. We then
repeated with the 8.4~GHz CBI simulated observations the same
deconvolution and mosaicking routines applied on the 31~GHz
pointings. This procedure provides us with comparable 8.4 and 31~GHz
images, as shown in Figure \ref{fig:2clr}-b. We caution that the
synthetic beam derived from the CBI $uv$-coverage is $4'$ FWHM, while
the input 8.4~GHz beam is $3'$ FWHM, and the MockCBI-processed 8.4~GHz
images has a final resolution of $5'$. Although the final resolution
of the observed and modelled images are comparable, non-random errors
are difficult to quantify (e.g. differences in the
deconvolution). These need to be considered when comparing flux
densities, particularly on the derived spectral indexes.

In order to include another point for the spectral study of Vela X, we
obtained from the Skyview Virtual Observatory\footnote{\tt
http://skyview.gsfc.nasa.gov} an image of the Vela X region at
4.8~GHz, acquired with the Parkes Telescope for the PMN Southern
survey \citep{co93}. The original image had $5'$ resolution. We tested
the calibration of the 4.8~GHz image by computing the integrated flux
from two reference sources mapped by the same survey and located near
Vela X, the point-sources PMN J1508-8003 and PMN J0823-5010, and
comparing with the reference fluxes obtained from the ATCA PMN
followup calibrator list\footnote{\tt
ftp://ftp.atnf.csiro.au/pub/data/pmn/CA/table2.txt}.  We reproduced
the MockCBI, MEM reconstruction and mosaicking procedures applied to
the 8.4~GHz data. The resulting 4.8~GHz image is shown in Figure
\ref{fig:2clr}-c , and has a final resolution of $6.4'$. We note that
the image from the PMN survey had missing pointings in Vela, but not
in that part of Vela X we are studying (the PWN and the brightest
filaments).

\subsection{Improvement of the 8.4~GHz image}

\begin{figure}[h!]
\begin{center}
\scalebox{0.9}[0.9]{\rotatebox{-90}
{\includegraphics{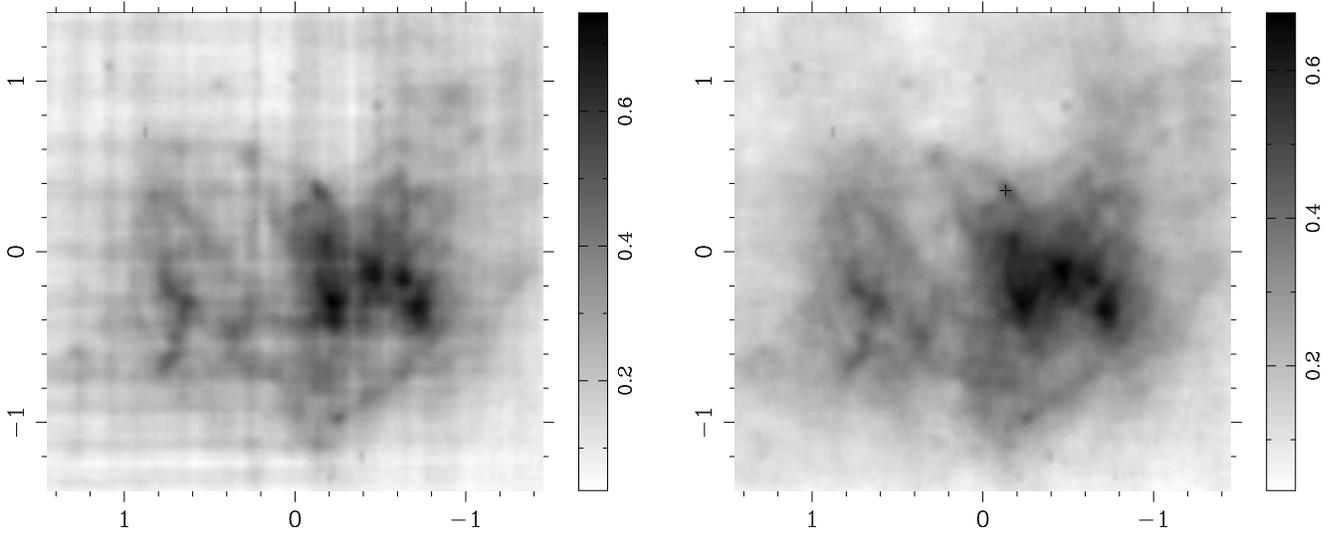}}} \caption{\small 8.4 GHz archive
image before and after destriping. The ($\alpha$, $\delta$) offsets
are with respect to (08:36:19.7,--45:23:38.4), units are in
degrees. Surface brightness units are in Jy~beam$^{-1}$, and the Parkes beam
FWHM is 3$'$. The position of the Vela pulsar is marked with a
cross.\label{fig:1clr} }
\end{center}
\end{figure}

Before simulating CBI observations over the 8.4~GHz data, we removed
several scanning artifacts that were present in the original 8.4~GHz
data. They appeared as many horizontal and vertical lines across the
sky image (Figure \ref{fig:1clr}-a). We removed these artifacts by
Fourier transforming the sky image and then suppressing contaminating
frequencies, which stand out as straight lines crossing the $(u,v)$
origin. We set to zero all visibilities whose moduli differ by more
than $3\sigma$ at a given $uv$-radius. Our destriping algorithm
follows a standard technique, detailed examples of which can be found
in \citet{em88}, \citet{da96}, and \citet{sc98}. The destriped 8.4~GHz
image is shown in Figure \ref{fig:1clr}-b.




\section{Results and Discussion}

\subsection{Vela X at 31~GHz}

\begin{figure}[h!]
\begin{center}
\scalebox{0.8}[0.8]{\rotatebox{-90}
{\includegraphics{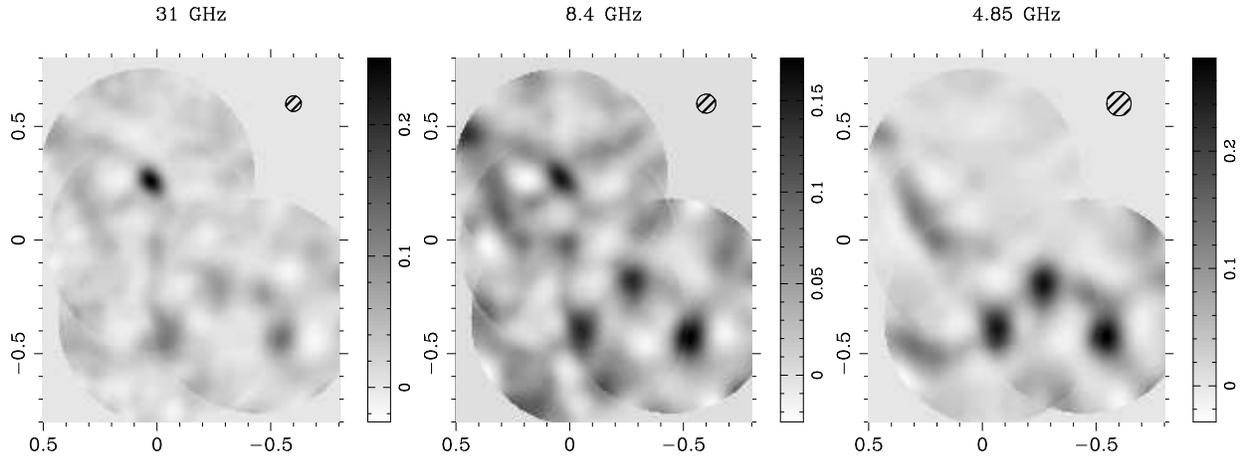}}
}
\caption{\small 31 GHz CBI observations of Vela X (a, left). 8.4 and
  4.8 GHz MockCBI-processed images are also presented (b and c, middle
  and right panels respectively). Map center is (08:35:10,--45:27:59),
  spatial units are in degrees and surface brightness is measured in
  Jy~beam$^{-1}$. The hatched ellipse drawn in the top right of each pannel
  represents each map's resolution (6.4$'$, 5$'$ and 4$'$
  respectively). \label{fig:2clr} }
\end{center}
\end{figure}

In Figure \ref{fig:2clr}-a, we present the 31-GHz CBI mosaic of Vela
X. We find a strong source around the Vela pulsar position, which we
associate with the radio PWN reported and described by
\citet{do03a}. We measure a FWHM size of $5.9'\times 4.1'$ in the final
MEM-deconvolved image. Alternatively, the model-fit decovolution
routine gives an estimated source size of $6.1'\times 4.1'$. The
orientation of the 31~GHz PWN is in agreement with that of the radio
lobes seen at higher resolution and at lower radio frequencies
\citep{le01,do03a}. This object represents the peak of the 31~GHz
emission (250~mJy~beam$^{-1}$), while filamentary structure is observed at
low emission levels. Figures \ref{fig:2clr}-b and \ref{fig:2clr}-c
show the mosaic images resulting from the simulation of the CBI's
$uv$-coverage on the 8.4-GHz and 4.8-GHz CBI reference images,
respectively. Comparison between the three images reveals a marked
brightening of the PWN at 31~GHz with respect to both 8.4-GHz and
4.8-GHz images, suggesting that the radio PWN has a flatter spectrum
than the rest of Vela X. The spectral analysis of our observations is
centered on the comparison between the PWN and the brightest regions
of the filaments.

The object we see at the pulsar position is not the Vela pulsar
itself. The pulsar is the brightest point source in four 12h
observations at 8.4~GHz made with ATCA, at 0.1~arcsec resolution (see
\citet{do03a} for more details), with a time-averaged flux density of
32$\pm$12~mJy (Dodson, private communication).  This is 1/10 our
lowest flux estimate for the PWN and the object we see is not
point-like but partially resolved by the CBI 4.1$'\times 4.1'$
resolution. Thus we ignored the Vela pulsar contribution in our
measures of the PWN fluxes. We also neglected the contamination due to
a background radio source near the Vela pulsar at 4.8, 8.4, and 31~GHz
since it´s 8.4~GHz integrated flux amounts to 7~mJy, or $1/100$ the
total PWN flux at 31~GHz \citep{do03a}.

\subsection{Fluxes and spectral energy distributions}


\subsubsection{Flux extraction}  \label{sec:flux}

\begin{figure}[h!]
\begin{center}
\scalebox{0.8}[0.8]{\rotatebox{-90} {\includegraphics{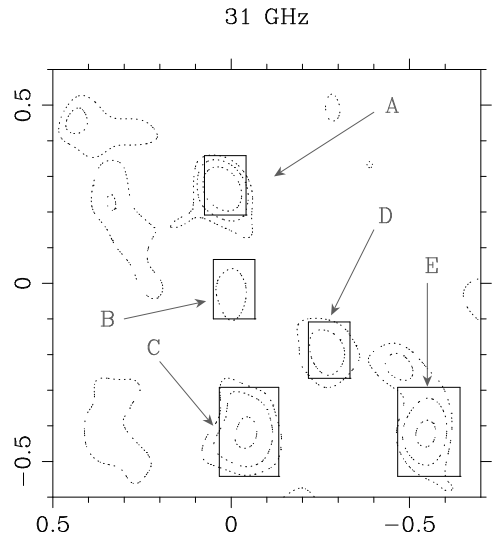}}}
\caption{\small Contour map of Vela X at 31~GHz (30 and 50 percent of
  peak emission) showing the photometric rectangular apertures (which
  we also call regions) where integrated flux densities and spectral
  indexes were computed. They are centered at the peak of each
  region. \label{fig:3clr} } \end{center}
\end{figure}

We smoothed the 31~GHz image to match the resolution of the 8.4~GHz
MockCBI-processed one, and the resolution of the 28.5~GHz image
($3.9'$) to match the resolution of the 33.5~GHz one ($4.8'$).  We
then used a photometric routine for computing the integrated fluxes
and spectral indexes of the Vela X brighter components, the PWN and
the brighter components of filamentary structure. We measured the
integrated flux within the photometric boxes shown in Figure
\ref{fig:3clr}. The noise level $\sigma$ was taken as the rms
dispersion of pixel intensities in a region of the image that excludes
the source. The statistical errors due to noise on the extracted flux
densities were estimated by multiplying the noise level $\sigma$ in
Jy~beam$^{-1}$ by $\sqrt{N/N_\mathrm{beam}}$, where $N$ is the number of pixels
within the aperture (or within the FWHM of the fitted ellipse), and
$N_\mathrm{beam}$ is taken to represent the number of correlated pixels (those
that fall in the FWHM of the beam). This is equivalent to multiplying
the noise level in Jy~pixel$^{-1}$ by $\sqrt{N\,N_\mathrm{beam}}$.

%

\begin{figure}[h!]
\begin{center}
\scalebox{0.5}[0.5]{\rotatebox{-90}
{\includegraphics{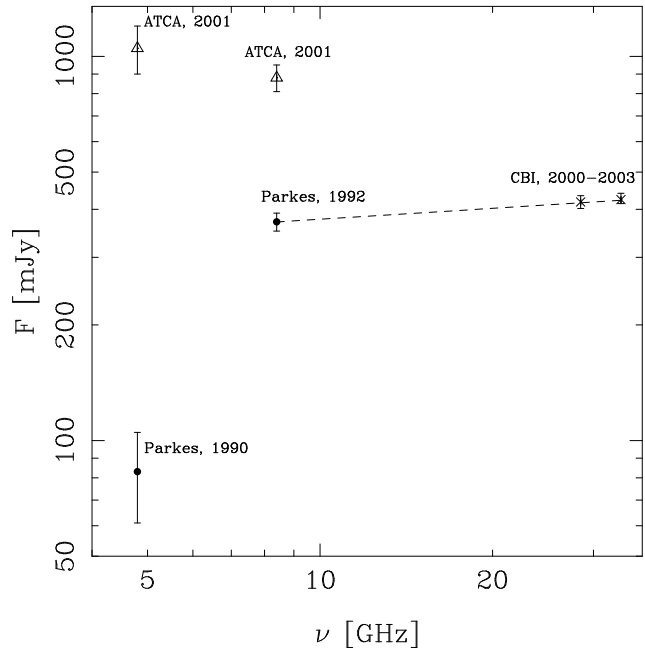}}
}
\caption{\small Integrated flux density spectrum for region A,
corresponding to the radio PWN. The straight line represents the
least-squares fit to the 8.4, 28.5 and 33.5~GHz data, given by
$\log{F}=(2.482\pm0.008)+(0.098\pm0.05)\log\nu$. The CBI and
CBI-simulated Parkes error bars represent statistical errors (see
Sec.~\ref{sec:flux}), while ATCA error bars are estimated from
variation in the definition of the regions \citep{do03a}. The scatter
at low frequencies is discussed in
Sec.~\ref{sec:var2}. \label{fig:4clr} }
\end{center}
\end{figure}

\begin{figure}[h!]
\begin{center}
\scalebox{0.5}[0.5]{\rotatebox{-90}
{\includegraphics{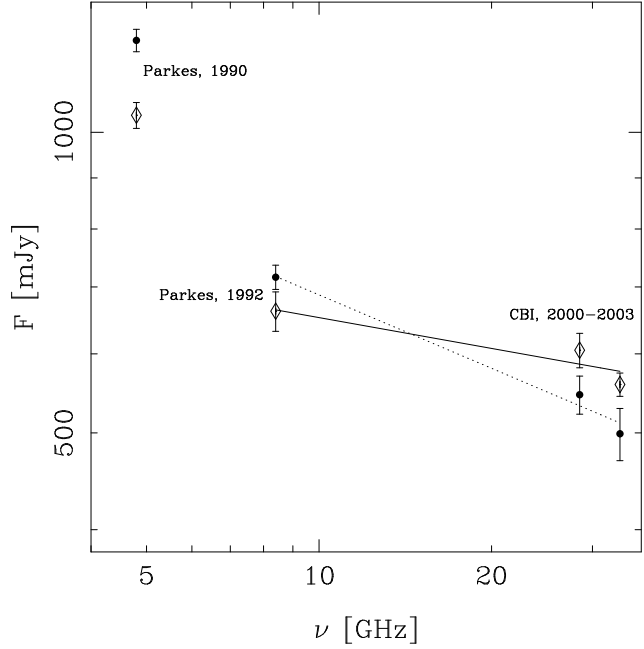}}
}
\caption{\small Integrated flux densities for region C (diamonds,
solid line) and region E (filled-circles, dashed line), corresponding
to the brightest regions of the filamentary structure. Errors are the
statistical estimates due to noise. The straight lines represents the
least-squares fits to each region's integrated flux density, which
gave spectral indexes of $\alpha_\mathrm{C}=-0.10\pm0.06$ and
$\alpha_\mathrm{E}=-0.24\pm0.04$, respectively. The fit does not
include the 4.8~GHz data. \label{fig:5clr} }
\end{center}
\end{figure}

The integrated fluxes measured within the photometric boxes (those
shown in Figure \ref{fig:3clr}), and their respective statistical
errors are presented in Table \ref{tbl-1}. The 31~GHz measurements are
not independent of 33.5~GHz and 28.5~GHz ones, but we quote all three
because they are used to compute the 8.4/31~GHz and the 28.5/33.5~GHz
spectral indexes separately. The fact that the 31~GHz measurements
shown in Table \ref{tbl-1} are not equal to the average of the
28.5~GHz and 33.5~GHz fluxes is due to the different resolutions of
the three maps. By smoothing the 31~GHz to the resolution of the
28.5~GHz image we obtain 31~GHz integrated fluxes consistent with the
28.5~GHz and 33.5~GHz measurements (In the case of the PWN, the flux
at 31~GHz changes from 411~mJy to 421~mJy when matching the 28~GHz
resolution). We have also included the 4.8~GHz integrated flux
values. In Figure~\ref{fig:4clr} we have plotted the flux density
spectrum for region A, corresponding to the Vela radio
PWN. Measurements from different epoch and instruments are also shown,
but a comparison between them is postponed to Section~\ref{sec:var2}.
In Figure~\ref{fig:5clr} we have plotted the flux density spectrum for
regions C and E, corresponding to the brightest regions of the Vela X
filamentary structure at 31~GHz.

At 31~GHz the PWN is approximately elliptical, and its flux can also
be obtained by a Gaussian fit. Instead of fitting an elliptical
Gaussian to the sky image, we preferred to work with the visibilities
on the $uv$-plane, providing us with an alternative (and
reconstruction independent) PWN integrated flux computation. Thus, for
the compact nebula surrounding the pulsar, we used the model-fitting
deconvolution routine implemented in the Difmap data reduction package
\citep{sh97}. The model-fitting routine essentially matches model
visibilities, calculated on a parametrized model image in the sky
plane, to the observed ones.  We used an elliptical Gaussian component
to model the compact radio nebula, whose physical parameters
(integrated flux, radial distance, position angle of the center of the
component, major axis, axial ratio and position angle of the major
axis) are adjusted by the model-fitting routine.

Thus, as an alternative to the photometry flux measurement, we show in
Table~\ref{tbl-2} the integrated flux densities given by the
model-fitting deconvolution routine. At 8.4~GHz the integrated flux
provided by model-fitting does not represent a good estimate for the
PWN, because at 8.4~GHz the modeled ellipse is much larger
($13.1'\times 4.5'$) than at 31-GHz. This is not related to the
frequency-size relationship observed in sources with power-law spatial
emission profiles \citep[for instance]{rei95}, derived from varying
beam sizes with frequency, because in our case the 8.4~GHz and 31~GHz
beams are very similar (and would be identical if the 8.4~GHz data had
a resolution much finer than its 3~arcmin).  We doubt the factor of
$\sim 2.5$ increase in major axis from 31~GHz to 8.4~GHz is a real
feature of the PWN. Rather, it is probably just that the extended
emission present at 8.4~GHz contaminates the model-fitting routine. We
thus constrained the morphological parameters of the 8.4~GHz
model-fitted ellipse in order to match those of the 31-GHz one,
leaving just the position and total flux variable. The latter gives a
8.4~GHz flux density value consistent with the one obtained with the
photometric routine in a box that approximately isolates the PWN
itself: $332\pm30$~mJy~beam$^{-1}$ for the size-constrained ellipse compared
to $371\pm20$~mJy~beam$^{-1}$, the flux density value obtained with the
photometric routine. We note that the centroid of the 8.4~GHz
constrained ellipse is slightly shifted NE from the 31~GHz one (towards
the North-East part of the PWN). This could be related to the steep
spectral index measured by \citet{bi91} for the North-Eastern lobe,
and could explain the large extension of the 8.4~GHz model-fitted
ellipse. The discussion on possible morphological changes between
8.4~GHz and 31~GHz is postponed to Section~\ref{sec:mor}. At 4.8~GHz the
model-fitting routine does not converge.

\subsubsection{Spectral indexes}

\begin{figure}[h!]
\begin{center}
\scalebox{0.8}[0.8]{\rotatebox{-90} {\includegraphics{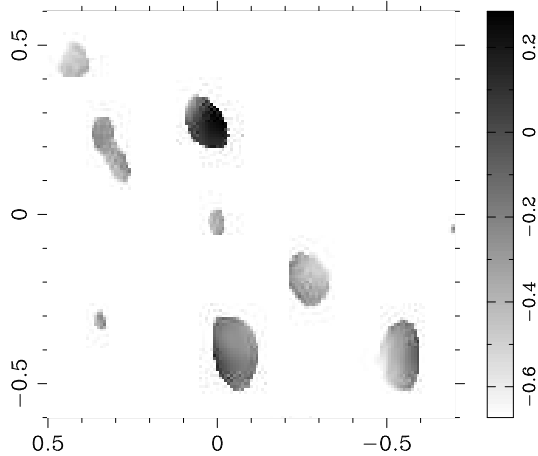}}}
\caption{\small Vela X 8.4/31~GHz spectral index distribution ($5'$
beam size). When computing the spectral index image we used a cutoff
level of 71~mJy and 51~mJy for the 8.4 and 31~GHz images,
respectively. Axis units are in degrees and the  center position is as for
Fig.~\ref{fig:2clr}. \label{fig:6clr} }
\end{center}
\end{figure}

Using the flux values presented in the previous sub-section, we
proceeded to compute the Vela X PWN and filaments spectral indexes.
We estimated the effects of varying the size of the photometric boxes
on the derived spectral indexes, keeping their centers fixed.  The
systematic uncertainty derived from the box definition dominates over
the statistical errors (those coming from flux calculation), and are
the ones shown in Table~\ref{tbl-3}, where we present spectral indexes
for different regions on the studied frequency bands. Our results
reveal that the PWN $\alpha_{8.4}^{31}$ spectral index is positive,
with a value of $\alpha_{8.4}^{31} = 0.10\pm0.06$, while the regions
of the filamentary structure have an almost uniform negative spectral
index, with an average of $\alpha_{8.4}^{31} = -0.28\pm0.09$ (in which
the uncertainty comes from the systematic error linked to the box
definition, and from variations between different regions). These
results are in agreement with the 8.4/31~GHz spectral index map shown
in Figure~\ref{fig:6clr}.

The $\alpha_{4.8}^{31}$ and the $\alpha_{28.5}^{33.5}$ spectral
indexes were computed after smoothing the higher frequency images to
the resolution of the lower frequency ones. The derived spectral
indexes are shown in Table~\ref{tbl-3}. The results obtained show
similar features to those arising from the 8.4/31~GHz comparison:
While the SED of the filaments follows a negative power-law, the PWN's
SED is flat (or even positive). The $\alpha_{8.4}^{31}$ spectral index
derived from the model-fitting routine and from the photometric boxes
are both positive, but differ at $\sim 2\sigma$. We used the
constrained ellipse for the calculation of the 8.4~GHz flux with the
model-fitting routine, as we believe it gives better account of the
PWN parameters than the unconstrained ellipse. It can be concluded
$\alpha_{8.4}^{31}>0.10\pm0.06$ for the PWN.

The $\alpha_{28.5}^{33.5}$ spectral index is also sensitive to the
flux extraction method. But at 26-36~GHz the PWN is very well fit by
an elliptical Gaussian, whereas the MEM-restored image is affected by
significant negative residuals from the synthetic beam
side-lobes\footnote{The Cornwell-Evans MEM \citep{cor85} implemented
in AIPS++ neglects the side-lobes of the synthetic beam, which is a
poor approximation in the case of the CBI.}. Thus we prefer the
$\alpha_{28.5}^{33.5}$ as calculated with the model-fitting
routine. However, we chose to fit the 28.5 and 33.5~GHz PWN
visibilities with ellipses of constrained size (having the size of the
best-fit ellipse for all 10 CBI channels) - unconstraining the
ellipses results in a steeper index of
$\alpha_{28.5}^{33.5}=-0.45\pm0.16$.  We conclude that
$\alpha_{28.5}^{33.5}<-0.25\pm0.16$.

In summary, $\alpha_{8.4}^{31}$ and $\alpha_{28.5}^{33.5}$ differ at
2$\sigma$, so we suspect a possible spectral break somewhere within
8.4-31~GHz in the SED of the PWN. However, a caveat must be made in
the sense that our flux estimations do not take into account
non-random errors due to instrumental effects or introduced by
differences in the deconvolution. Biases in the total flux densities
in MEM deconvolved maps are well known \citep{cor99}, as
Cornwell-Evans MEM needs the total flux parameter in their approximate
MEM optimization. Substantial additional flux has been reported both
in simulations and in real observations. These effects may affect our
spectral index estimations. In consequence, we caution that the
possible spectral break somewhere within 8.4-31~GHz should be
considered just as a marginal detection, since it could easily be an
instrumental effect. However, our main result still holds: the PWN has
a flatter spectral index than the rest of Vela X, as meets the eye in
Fig.~\ref{fig:2clr}.




\subsection{Nature of the Vela X PWN and filamentary structure} \label{sec:var}

The relatively flat SED of the PWN contrasts with the steeper spectrum
of the filamentary structure, as clearly seen in
Figures~\ref{fig:4clr} and \ref{fig:5clr}. These differences suggest
that the PWN is differently, or more recently, energized by the
pulsar. Flat spectrum cores are characteristic of pulsar-powered
nebulae \citep{rei02} and are predicted by theoretical models of
pulsar winds \citep{rey03}. It is possible that the Vela PWN has a
flat SED between 8.4 and 31~GHz, and was barely seen at lower radio
frequencies since it was hidden by the steep spectrum emission of the
radio filaments and diffuse emission. However, it is also possible
that the flat PWN SED detected arises from recent particle
injection. In Section~\ref{sec:var2} We investigate on the
variability hypothesis.

\begin{figure}[h!]
\begin{center}
\scalebox{0.6}[0.6]{\rotatebox{-90} {\includegraphics{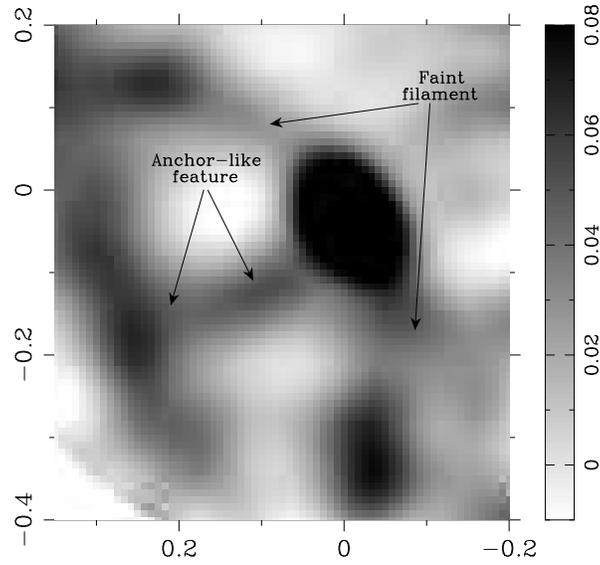}}}
\caption{\small PWN surroundings at 31~GHz. The faint filament
extending SW from the PWN and the anchor-like feature are clearly
seen. Axis units are in degrees and map center is
(08:35:20,--45:10:20).\label{fig:7clr} }
\end{center}
\end{figure}

Most authors have agreed on the plerionic nature of Vela X
\citep{dw91,al01,bo98}, but the relationship between the pulsar and
the rest of Vela X is still unclear. Our results should provide key
pieces in the puzzle of this relationship:
\begin{itemize}
\item The spectral indexes we obtain for the filamentary structure are
in good agreement with the ones published by \citet{al01} for the
whole of Vela X. \citet{mi95} found no evidence for steepening of the
spectrum with increasing distance from pulsar, as would be expected in
pulsar-powered remnants. However, our results show that the region
surrounding the pulsar does have a flatter SED than the radio
filaments.

\item In our 31~GHz observations (Figure \ref{fig:2clr}-a), running
southwards from the PWN we see the one-sided southern `jet' starting
SW from the PWN and then turning South from the PWN, crossing two
regions of enhanced emission (regions B and C in
Fig.~\ref{fig:3clr}). From Figure \ref{fig:7clr}, in which we have
zoomed the PWN surroundings, it is apparent that the PWN is linked to
the southern `jet' by a faint filament that seems to be an extension
of the PWN's SW lobe. This filament also stands out as a high
spatial-frequency structure at 8.4~GHz, as can be seen in the
MockCBI-simulated image in Figure \ref{fig:2clr}-b. The possibility of
a chance projection of filamentary structure over the PWN is unlikely,
because the tangent of the filament at the PWN's position is parallel
to the PWN's major axis. To further back the above points suggesting
pulsar-powered filaments, we note another filament is seen running SE
from the PWN, and ending in an `anchor-like' feature, which is located
in the direction of proper motion and spin axis of the pulsar. These
features are present the MOST mosaic of the Vela SNR by \citet{bo98} ,
although they are more evident in our 31~GHz CBI images. Objects such
as G11.2-0.3, G18.95-1.1 and G21.5-0.9, cataloged as plerions and
combined-type SNRs, exhibit similar characteristics as the ones
described above for Vela X: They exhibit filaments running outwards
from a central bar-like feature \citep{rei02}. The central feature in
these plerions is thought to be responsible for the particle injection
process to the filaments.

\item The regions we see as filaments are the ones described by Milne
as `polarized radio filaments' (\citet{mi95}, his Figure 2-b), with
the magnetic field directed along these filaments, which he notes is
`A feature previously seen only in the Crab Nebula \citep{ve92}', and
also emphasizes their similar degree of polarization. In Milne's
polarized intensity maps, the PWN is seen at similar polarization
fractions than the filamentary structure, and almost completely hidden
by diffuse emission in total intensity.
\end{itemize}

Our observations support aspects of the models by \citet{rey88} and
\citet{bl01}. \citet{rey88} proposed an explanation for filamentary
structure in Crab-like SNRs, consisting of thermal filamentary
structure (caused by Rayleigh-Taylor instabilities operating on
thermal gas accelerated by the pulsar's wind), interacting with
pulsar-generated relativistic particles. Variation in the radio
spectrum of the filaments is expected from Reynold's model, an effect
we do not see in Vela X along the filaments, but rather between the
radio filaments and the PWN surrounding the pulsar. \citet{bl01} have
proposed a model for Vela X in which filamentary structure is not
directly pulsar-powered, but rather originated from the
Rayleigh-Taylor instability during the crushing of the original PWN by
a reverse shock wave that can give account of the filamentary
structure's chaotic appearance and off-center pulsar position. However
the model by \citet{bl01} does not address the PWN, and does not
explain the spectral properties of the PWN presented in this work and
its links with the filamentary structure.

\subsection{Morphological changes between the 8.4~GHz and 31~GHz images}  \label{sec:mor}

\begin{figure}[h!]
\begin{center}
\scalebox{0.6}[0.6]{\rotatebox{-90}
{\includegraphics{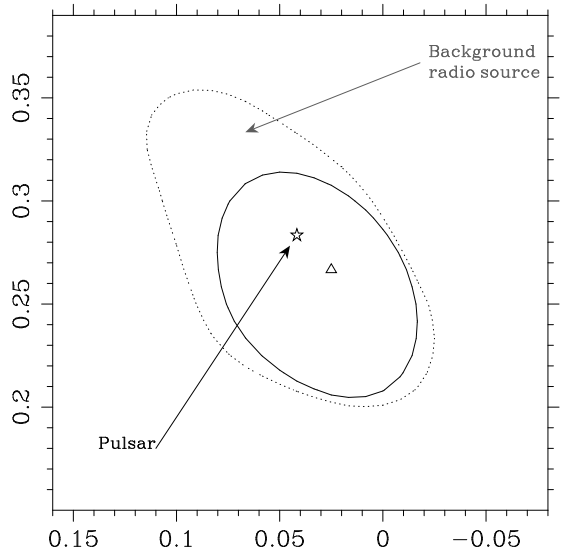}}} \caption{\small 8.4 GHz
(dotted) and 31 GHz (light grey solid) contours at half-maximum,
together with the position of the PWN emission maxima (star: 8.4~GHz,
triangle: 31~GHz). Axis units are in degrees and
center position is as for Fig.~\ref{fig:2clr}.\label{fig:8clr} }
\end{center}
\end{figure}

As previously mentioned at the end of Section~\ref{sec:flux}, the
center of the 8.4~GHz and 31~GHz fitted ellipses are offset from each
other. This effect is also seen when comparing the position of each
image maxima: While at 8.4~GHz the peak of emission approximately
coincides with the pulsar's position, at 31~GHz the maximum emission
is shifted in the SW direction. This shift is evident in
Figure~\ref{fig:8clr}, where we show the position of the PWN emission
maxima at 8.4 and 31~GHz, together with the respective contours at
half-maximum. We estimated the pointing accuracy of the CBI by
analyzing phase calibrator observations, obtaining a $20''$~rms
pointing accuracy for the Vela observations. The shift between both
maxima is thus $85''\pm20''$. The shift between the model-fitted
ellipse centers (constraining the ellipse size at 8.4~GHz to match
that at 31~GHz) is $78''\pm20''$.

\begin{figure}[h!]
\begin{center}
\scalebox{0.6}[0.6]{\rotatebox{-90}
{\includegraphics{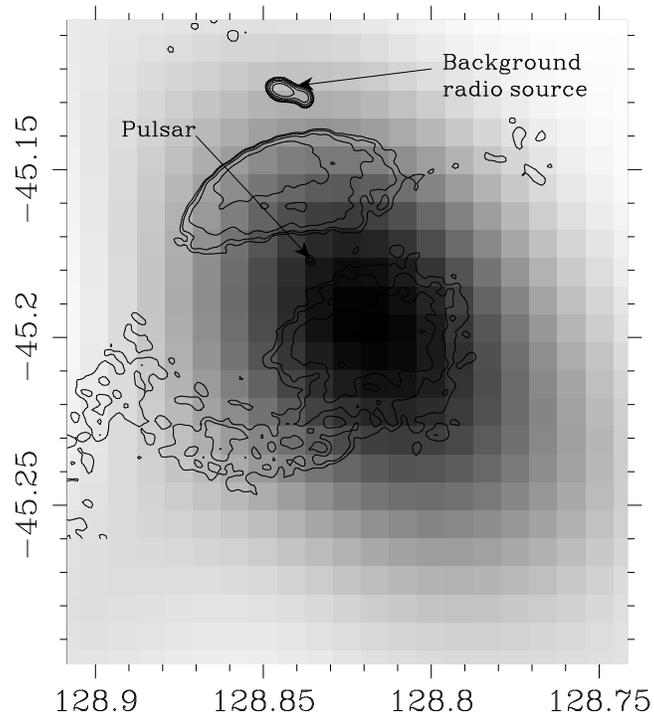}}} \caption{\small Compact radio
source at 5~GHz (contours, ATCA data, \citet{do03a}) and 31 GHz
(greyscale). The Vela pulsar and the background radio source are
clearly seen. \label{fig:9clr} }
\end{center}
\end{figure}

In Figure~\ref{fig:9clr} we have overlaid a 5 arcsec resolution ATCA
map at 5~GHz \citep{do03a} in contours on the 31~GHz CBI PWN
observations in grey scale. In spite of the resolution differences,
the shift between the peaks of emission is apparent: it lies in the NE
lobe in the ATCA data, while it is closer to the SW lobe in the CBI
data. This effect is also seen in the 8/31~GHz spectral index
distribution (Figure~\ref{fig:6clr}), as a steepening of the PWN
spectrum northeastwards. Indeed, this steepening is in agreement with
the scaled array VLA observations of \citet{bi91}, who previously
reported the steep spectral index of the NE feature, but did not
detect the `flat spectrum' emission from the SW lobe. The spectral
index variations detected within the PWN predict morphological changes
with increasing frequency.

\subsection{Investigation on possible PWN variability} \label{sec:var2}

Could the brightening of the region surrounding the Vela pulsar in our
31~GHz data result from variability of the PWN ? This effect have not
been reported yet at radio-frequencies, but already evidenced in
X-rays by the recent Chandra images of the variable PWN X-ray
structure \citep{pav03}.

PSR B0833$-$45, the Vela pulsar, is characterized by giant glitches,
or sudden spin-ups, and is the pulsar where such deviations from
steady slow down were first discovered. The latest glitch in the Vela
pulsar, which is also the largest spin-up among all glitching pulsars,
was recorded in January 2000 \citep{do00}. As pulsars are the main
responsibles for the particle injection in pulsar-powered nebula
models, a reasonable question arises : Do giant glitches result in
visible changes in the surrounding pulsar-powered nebula ? We suspect
variability of the PWN since the integrated flux that we obtain for
the PWN at 4.8 and 8.4~GHz from the MockCBI-processed images are well
below the values published by \citet{do03a}, as may be inferred from
Figure~\ref{fig:4clr}, where we have plotted ATCA 2001 PWN fluxes,
obtained by adding the fluxes from the SW and NE lobes of the PWN
published by \citet{do03a}. Observation with ATCA in 2001 measure a
higher 4.85~GHz flux density than that from the Parkes 1990
image. This increase of the PWN flux between both epochs is probably
not merely due to different instrument responses.  The difference is opposite
to that expected from the ATCA and Parkes responses: because of
incomplete sampling in the $uv$ plane the  interferometer data is
insensitive to small spatial frequencies.  However, to perform a
reliable flux comparison the single-dish observations should have been
modeled onto the interferometer which is being compared (as done for
the 8.4/31~GHz data).

The flux density differences may also be influenced by the poor
sensitivity of the PMN survey to low spatial-frequency
structures. \citet{co93} explain that sources extended more than 30$'$
in declination are supressed in the PMN survey. But such spatial
frequency corresponds to a $uv-$radius of 115~rad$^{-1}$, which is
similar the minimum baseline length for the CBI, of
90~rad$^{-1}$. Thus a Parkes-CBI comparison based on CBI-simulated
images should be accurate at all observed angular scales. The
Parkes/CBI comparison of the PWN flux is independent of missing
angular frequencies in the PMN data because it is much smaller than
the problematic angular scales.

Our CBI observations of the PWN in November 2000 and April 2003,
following the Vela pulsar's January 2000 glitch, showed no evidence of
variation of the PWN integrated flux density ($389\pm8$~Jy and
$406\pm11$~Jy for the 2000 and 2003 observations
respectively\footnote{The April 2003 observations were carried out
using a more compact antenna configuration than of the November 2000
observation. Thus to compare both epochs we restricted the range in
the $uv$-radius from 150 to 400 $\lambda$, which is the range covered
by the 2003 observations.}). It is likely that the differences
measured between the Parkes and ATCA observations are due to
instrumental/imaging effects overall. Indeed, \citet{bi91} did not
find any change in the NE feature within the four-months interval of
their observations.  Our analysis only allows us to conclude that any
possible change in the pulsar surroundings must have been produced
before our observation epoch, 9 months after the strong January 2000
glitch. This conclusion states that the total flux within a
($5.9'\times\,4.1'$) elliptical region has varied. Variations in the
extension of the region could also be expected, but at the distance of
the Vela pulsar (d~$287$ pc, as derived from recent VLBI observation
by \citet{do03b}) the $5.9'$ FWHM size of the PWN is just about the
radius covered by light in 9 months, and is hence unlikely to be
detected within the epoch of the CBI observations. It is clear that
only high spatial-resolution time-monitoring of the pulsar surrounding
can provide reliable conclusions on possible time-dependent changes
within the radio PWN.

\section{Conclusions}

We have detected a strong radio source at 31~GHz around the Vela
pulsar, that we identify with the radio pulsar wind nebula (PWN)
reported by \citet{do03a}. The PWN is surrounded by a low-level
network of filaments. We report dramatic changes with frequency in the
morphology of Vela X, the pulsar vicinity becoming markedly brighter
relative to the rest of Vela X as the frequency increases from 4.8~GHz
to 31~GHz.  The spectrum of the radio PWN is flat, with an spectral
index value of $\alpha_{8.4}^{31}=0.10\pm0.06$. In contrast, the
spectral index of the Vela X filaments is negative, with an average
value of $\alpha_{8.4}^{31}=-0.28\pm0.09$. We investigate whether the
flat spectrum PWN obtained in this work reflects the SED of the object
or is a consequence of variability (or a combination of both effects),
but the lack of consistent comparison data does not allow firm
conclusions to be drawn on this subject.

As a step towards understanding the nature of the filamentary
structure, we observe that a faint filament crosses the PWN, with a local
tangent at the position of the PWN sharing the same position angle as
the PWN major axis. This feature might be associated with an extension
of the PWN itself. A pulsar-powered nature for the Vela X filaments is
also supported by the similarity in morphology and polarization with
other plerionic cores.

We detect a shift in the low-resolution centroid of the PWN between
8.4 and 31~GHz, by $80''\pm20''$, revealing variations of the spectral
index within the PWN. The shift is also evident when comparing the
31-GHz CBI data with high resolution 5-GHz images.

\acknowledgments

We are very grateful to Doug Milne and Patricia Reich for making
available the 8.4~GHz image, and to Doug Bock for providing his
0.843~GHz image.  S.C acknowledges support from Fondecyt grant
1030805.  S.C., J.M., and L.B. acknowledge support from the Chilean
Center for Astrophysics FONDAP 15010003. We gratefully acknowledge the
generous support of Maxine and Ronald Linde, Cecil and Sally
Drinkward, Barbara and Stanley Rawn, Jr., and Fred Kavli.  This work
is supported by the National Science Foundation under grant AST
00-98734.

\clearpage

\begin{deluxetable}{crrrrr}
\tablecaption{ Vela X integrated flux densities (mJy). \label{tbl-1}}
\tablewidth{0pt} \tablehead{ \colhead{Frequency (GHz)} &
\colhead{Region A$^{\dag}$}& \colhead{Region B} & \colhead{Region C} &
\colhead{Region D} & \colhead{Region E} } \startdata 33.5 & 424$\pm$10
& 262$\pm$20 & 559 $\pm$15 &232 $\pm$15 & 499 $\pm$30 \\ 31 &
411$\pm$10 & 261$\pm$30 & 552 $\pm$15 &226 $\pm$15 & 493 $\pm$30 \\
28.5 & 418$\pm$16 & 285$\pm$24 & 605 $\pm$24 &237 $\pm$15 & 546
$\pm$24\\ 8.4 & 371$\pm$20 & 360$\pm$20 & 662 $\pm$30 &375 $\pm$20 &
716 $\pm$20\\ 4.85 & 83$\pm$22 & 248$\pm$20 & 1040 $\pm$31 &963
$\pm$22 & 1236 $\pm$32\\
\enddata
\tablenotetext{\dag}{the PWN}
\end{deluxetable}

\begin{deluxetable}{crrr}
\tablecaption{PWN flux densities from model-fitting. \label{tbl-2}}
\tablewidth{0pt} \tablehead{ \colhead{Frequency (GHz)} &
\colhead{Flux density (mJy)} }
\startdata 33.5  &529$\pm$40 \\
             31  &543$\pm$30 \\
            28.5 &551$\pm$40 \\
             8.4 &702$\pm$50 \\
 8.4 constrained &332$\pm$30 \\
\enddata
\end{deluxetable}


\begin{deluxetable}{crrrrr}
\tablecaption{Vela X spectral indexes. \label{tbl-3}}
\tablewidth{0pt} \tablehead{ \colhead{Component} &
\colhead{$\alpha_{8.4}^{31}$} & \colhead{$\alpha_{4.8}^{31}$} &
\colhead{$\alpha_{28.5}^{33.5}$} } \startdata Region A &$
$0.10$\pm0.06$ &$ $ 0.75$\pm0.05$ & $ $ 0.13$\pm0.11$ \\
model-fit$^{\dag}$&$ $0.37$\pm0.11$ & & $-$ 0.25$\pm0.16$ \\

Region B&$-$0.28$\pm0.12$ &$-$ 0.06$\pm0.11$ &  $-$ 0.43$\pm0.15$ \\
Region C &$-$0.16$\pm0.09$ &$-$ 0.41$\pm0.04$ & $-$ 0.35$\pm0.12$ \\
Region D &$-$0.39$\pm0.11$ &$-$ 0.65$\pm0.09$ & $-$ 0.14$\pm0.13$ \\
Region E &$-$0.24$\pm0.09$ &$-$ 0.54$\pm0.03$ & $-$ 0.28$\pm0.21$ \enddata

\tablenotetext{\dag}{For the PWN (photometric region A). The model-fit
spectral indexes are the ones computed by using the integrated flux
densities obtained by fitting an elliptical gaussian to the PWN.}
\end{deluxetable}


\clearpage

%






\begin{thebibliography}{}

\bibitem[Alvarez et al.(2001)]{al01} Alvarez, H., Aparici, J., May, J. \& Reich, P., 2001, \aap, 372, 636
\bibitem[Bietenholz, Frail \& Hankins(1991)]{bi91} Bietenholz, M.F., Frail, D.A., Hankins, T.H., 1991, \apj, 376; L41-L44
\bibitem[Blondin et al.(2001)]{bl01} Blondin, J., Chevalier, R.  \& Frierson, D., 2001, \apj, 563, 806
\bibitem[Bock et al.(2002)]{bo02} Bock, D.C.J., Sault, R.J., Milne, D.K. \& Green, A.J., 2002, ASP Conference series, Vol 271, 187
\bibitem[Bock et al.(1998)]{bo98} Bock, D.C.J., Turtle, A.J. \& Green, A.J., 1998, \aj, 116, 1886
\bibitem[Condon el al.(1993)]{co93} Condon, J.J., Griffith, M.R. \& Wright, A.E., 1993, \aj, 106, 1095
\bibitem[Cornwell, Braun \& Briggs (1999)]{cor99} Cornwell, T.J., Braun, R., Briggs, D.S., 1999, ASP Conf. Ser. 180, 151
\bibitem[Cornwell \& Evans (1985)]{cor85} Cornwell, T.J. \& Evans, K.F., 1985, \aap, 143, 77
\bibitem[Chevalier(2003)]{ch03} Chevalier R.A., 2003, in "High Energy
Studies of Supernova Remnants and Neutron Stars" (COSPAR 2002),
Advances in Space Research, {\em in press}, astro-ph/0301370
\bibitem[Davies et al.(1996)]{da96} Davies, R.D., Watson, R.A. \& Guti\'errez, C.M, 1996, \mnras, 278, 925
\bibitem[Dodson et al.(2000)]{do00} Dodson, R.G., McCulloch, P. M. \& Costa, M. E., 2000, IAU Circ., 7347, 2, Edited by Green, D. W. E. 
\bibitem[Dodson et al.(2003)]{do03a} Dodson, R.G., Lewis, D., McConnell, D. \& Deshpande, A.A., 2003, \mnras, 343, 116
\bibitem[Dodson et al.(2003b)]{do03b} Dodson, R.G., Legge, D.,
Reynolds, J.E., and McCulloch, P., M., 2003, \apj, 596, 1137
\bibitem[Dwarakanath(1991)]{dw91} Dwarakanath, K.S., 1991, J.Astrophys.Astr., 12, 199
\bibitem[Emerson \& Gr\"ave(1988)]{em88} Emerson, D.T. \& Gr\"ave, R.,
  1988, \aap,  190, 353
\bibitem[Frail \& Scharringhausen(1997)]{fr97} Frail, D.A. \& Scharringhausen, B.R. , 1997, \apj , 480, 364
\bibitem[Frail et al.(1997)]{fr97b} Frail, D.A., Bietenholz, M.F., Markwardt, C.B. \& \"{O}gelman, H., 1997, \apj, 475, 224
\bibitem[Green(2004)]{gr04} Green, D.A., 2004, ``A Catalogue of
Galactic Supernova Remnants'', Mullard Radio Astronomy Observatory,
Cambridge, United Kingdom
\bibitem[Helfand et al.(2001)]{he01} Helfand, D., Gotthelf, E. \& Halperm, J.P. , 2001, \apj, 556, 380
\bibitem[Kennel \& Coroniti(1984)]{ke84} Kennel, C.F. \& Coroniti, F.V., 1984, \apj, 283, 710
\bibitem[Lewis et al.(2002)]{le01} Lewis, D., Dodson, R.,
McConnell, D. \& Deshpande, A., 2002, in ASP Conf.Ser. Vol 271, Neutron Stars in Supernova Remnants, edited by Slane, P.O \& Gaensler, B.M. (San Francisco:ASP), 191
\bibitem[Markwardt \& \"{O}gelman(1995)]{ma95}  Markwardt, C.B. \& \"{O}gelman, H., 1995, \nat, 375, 40
\bibitem[Mason et al.(2003)]{ma03} Mason, B. S. , Pearson, T. J.,
 Readhead, A. C. S. , Shepherd, M.C.  Sievers, J. L., Udomprasert,
 P. S., Cartwright, J. K., Farmer, A. J., Padin, S., Myers, S. T.,
 Bond, J. R., Contaldi, C. R., Pen, U.-L., Prunet, S., Pogosyan, D.,
 Carlstrom, J. E., Kovac, J., Leitch, E. M., Pryke, C., Halverson,
 N. W., Holzapfel, W. L., Altamirano, P., Bronfman, L., Casassus, S.,
 May, J. \& Joy, M., 2003, \apj, 591, 540
\bibitem[Milne(1995)]{mi95} Milne D.K, 1995, \mnras, 277, 1435
\bibitem[Milne \& Manchester(1986)]{mi86} Milne, D.K \& Manchester R.N., 1986, \aap, 167, 117
\bibitem[Padin et al.(2002)]{pa02} Padin, S., Shepherd, M.C.j
Cartwright, J. K., Keeney, R. G., Mason, B. S., Pearson, T. J.,
Readhead, A. C. S., Schaal, W. A., Sievers, J., Udomprasert, P. S.,
Yamasaki, J. K., Holzapfel, W. L., Carlstrom, J. E., Joy, M., Myers,
S. T. \& Otarola, A. , 2002, PASP, 114, 83
\bibitem[Pavlov et al.(2001a)]{pav01a} Pavlov, G.G., Zavlin, V.E., Sanwal, D., Burwitz, V. \& Garmire, G.P., 2001, \apj, 552, L129
\bibitem[Pavlov et al.(2001b)]{pav01b} Pavlov, G.G., Kargaltsev, O.Y., Sanwal, D. \& Garmire, G.P., 2001, \apj, 554, L189
\bibitem[Pavlov et al.(2003)]{pav03} Pavlov, G.G., Teter, M.A., Kargaltsev, O.Y. \& Sanwal, D., 2003, \apj, 591, 1157
\bibitem[Radhakrishnan \& Deshpande(2001)]{ra01} Radhakrishnan, V. \& Deshpande A. , 2001, \aap, 379, 551
\bibitem[Rees \& Gunn(1974)]{re74} Rees, M.J. \& Gunn, J.E., 1974, \mnras, 167, 1-12
\bibitem[Reich et al.(1998)]{rei98} Reich W., F\"{u}rst, E. \& Kothes, R., 1998, Memorie della Societa Astronomia Italiana, Vol. 69, p.933
\bibitem[Reich(2002)]{rei02} Reich W., 2002, proceeding of the 270
``WE-Heraeus Seminar on Neutron Stars, Pulsars, and Supernova
Remnants.'', ed. Becker, W., Lesch, H., Tr\"{u}mper, J.
\bibitem[Reid et al.(1995)]{rei95} {Reid}, M.J.,  {Argon}, A.L.,
    {Masson}, C.R. and {Menten}, K.M. \& {Moran}, J.M., 1995, {\apj},
    443, {238}
\bibitem[Reynolds \& Chevalier(1984)]{rey84} Reynolds S.P. \& Chevalier R.A., 1984, \apj, 555, L49
\bibitem[Reynolds(1988)]{rey88} Reynolds S.P., 1988, \apj, 327, 853R
\bibitem[Reynolds(2003)]{rey03} Reynolds S.P., 2003,  Proceedings of
  IAU Colloquium 192 ``10 Years of SN1993J'', {\em in press}, astro-ph/0308483
\bibitem[Schlegel et al. (1998)]{sc98} Schlegel D.J., Finkbeiner D.P. \& Davis M., 1998, \apj, 500, 525
\bibitem[Shepherd(1997)]{sh97} Shepherd M.C., 1997, in Astronomical
Data Analysis Software and Systems VI, ed. G~Hunt \& H.E.~Payne, ASP
conference series, v125, 77 ``Difmap: an interactive program for
synthesis imaging''.
\bibitem[Velusamy et al.(1992)]{ve92} Velusamy, T., Roshi, D. \& Venugopal, V.R., 1992, \mnras, 255, 210
\bibitem[Weiler \& Panagia(1980)]{we80} Weiler, K.W. \& Panagia, N., 1980, \aap, 90, 269




%
%


\end{thebibliography}
\end{document}